\documentclass[aps,prd,onecolumn,showpacs,amsmath,amssymb,superscriptaddress,floatfix,nofootinbib,11pt]{revtex4-1}

\usepackage[colorlinks=true, citecolor=blue, linkcolor=blue, urlcolor=blue]{hyperref}
\usepackage{color,xcolor,graphicx,amsfonts,multirow,amsmath,multirow}
\usepackage{ulem}

\allowdisplaybreaks

\begin{document}

\title{The rare decay $B^+ \to K^+\ell^+\ell^-(\nu\bar{\nu})$ under the QCD sum rules approach}

\author{Hai-Jiang Tian}
\email{THUnsong@outlook.com}
\address{Department of Physics, Guizhou Minzu University, Guiyang 550025, P.R.China}

\author{Hai-Bing Fu}
\email{fuhb@gzmu.edu.cn}
\address{Department of Physics, Guizhou Minzu University, Guiyang 550025, P.R.China}
\address{Institute of High Energy Physics, Chinese Academy of Sciences, Beijing 100049, P.R.China}

\author{Tao Zhong}
\email{zhongtao1219@sina.com}
\address{Department of Physics, Guizhou Minzu University, Guiyang 550025, P.R.China}
\address{Institute of High Energy Physics, Chinese Academy of Sciences, Beijing 100049, P.R.China}

\author{Ya-Xiong Wang}
\email{wangyx12138@sina.com}
\address{Department of Physics, Guizhou Minzu University, Guiyang 550025, P.R.China}

\author{Xing-Gang Wu}
\email{wuxg@cqu.edu.cn}
\address{Department of Physics, Chongqing Key Laboratory for Strongly Coupled Physics, Chongqing University, Chongqing 401331, P.R.China}

\begin{abstract}

In the paper, we conduct a detailed investigation of the rare decay processes of charged meson, specifically $B^+ \to K^+\ell^+\ell^-$ with $\ell=(e,\mu,\tau)$ and $B^+ \to K^+\nu\bar{\nu}$. These processes involve flavor-changing-neutral-current (FCNC) transitions, namely $b\to s\ell^+\ell^-$ and $b\to s\nu\bar{\nu}$. The essential components $B\to K$ scalar, vector and tensor transition form factors (TFFs) are calculated by using the QCD light-cone sum rules approach up to next-to-leading order QCD corrections. In which, the kaon twist-2 and twist-3 light-cone distribution amplitudes are calculated from both the QCD sum rules within the framework of background field theory and the light-cone harmonic oscillator model. The TFFs at large recoil point are $f_+^{BK}(0)=f_0^{BK}(0) =0.328_{-0.028}^{+0.032}$ and $f_{\rm T}^{BK}(0)=0.277_{-0.024}^{+0.028}$, respectively. To achieve the behavior of those TFFs in the whole $q^2$-region, we extrapolate them by utilizing the simplified $z(q^2)$-series expansion. Furthermore, we compute the differential branching fractions with respect to the squared dilepton invariant mass for the two different decay channels and present the corresponding curves. Our predictions of total branching fraction are ${\cal B}(B^+\to K^+ e^+ e^-)=6.633_{-1.070}^{+1.341}\times 10^{-7}$, ${\cal B}(B^+\to K^+ \mu^+ \mu^-)=6.620_{-1.056}^{+1.323}\times 10^{-7}$, ${\cal B}(B^+\to K^+ \tau^+ \tau^-)=1.760_{-0.197}^{+0.241}\times 10^{-7}$, and ${\cal B}(B^+\to K^+ \nu\bar{\nu})=4.135_{-0.655}^{+0.820}\times 10^{-6}$, respectively. Lastly, the observables such as the lepton universality $\mathcal{R}_{K}$ and the angular distribution `flat term' $F_{\rm H}^\ell$ are given, which show good agreement with the theoretical and experimental predictions.

\end{abstract}

\date{\today}


\maketitle

\newpage

\section{Introduction}\label{Sec:1}

The Standard Model (SM) is one of the most successful theories, offering our most refined depiction of fundamental particles to date. From both experimental and theoretical perspectives, testing and challenging the SM, as well as searching for new physics beyond the SM, are highly significant. The flavor-changing-neutral-current (FCNC) transitions in the $B$-factories, such as $b\to s\ell^+\ell^-$ with $\ell=(e,\mu,\tau)$ and $b\to s\nu\bar{\nu}$, can provide such an opportunity. Meanwhile, the semi-leptonic $B$-meson decays are usually regarded as particularly valuable probes for testing the SM, as they exhibit unique and clean experimental signatures and have controllable theoretical uncertainties. In particular, the FCNC processes $b\to s\ell^+\ell^-$ proceed at loop level in the SM while its contributions are suppressed by loop factors and related elements of the Cabibbo-Kobayashi-Maskawa (CKM) matrix~\cite{Cabibbo:1963yz, Kobayashi:1973fv}, and $b\to s\nu\bar{\nu}$ also will be suppressed by the Glashow-Iliopoulos-Maiani (GIM) mechanism~\cite{Glashow:1970gm}.

The $B^+\to K^+\ell^+\ell^-$ and $B^+\to K^+\nu\bar{\nu}$ decays have recently garnered significant attention, particularly in light of the new results from the LHCb and the Belle Collaborations, inspiring our interest in exploring the properties of these two processes. Below are some brief descriptions of the current status with regard to these processes from both the experimental and theoretical sides. In 2014, the LHCb Collaboration conducted measurements of lepton universality (LU) in the $B^+\to K^+\ell^+\ell^-$ decay processes, with the aim of testing the SM~\cite{LHCb:2014vgu}. The results, which obtained within the dilepton mass squared range $q^2\in [1.0,6.0]~{\rm GeV^2}$, showed a value of ${\cal R}_K=0.745_{-0.074}^{+0.090}({\rm stat.})\pm0.036({\rm sys.})$. This value deviated from the SM value of unity by about 2.6 standard deviations. Therefore, the LHCb Collaboration continued to measure the LU during the ensuing 2019, 2021 and 2022 years~\cite{LHCb:2019hip, LHCb:2021trn, LHCb:2022vje}. Especially for the most recent LHCb measurement in 2022, by using nonresonant $B^+\to K^+\ell^+\ell^-$ and $B^+\to K^{*0}\ell^+\ell^-$ decays obtained $\mathcal{R}_K$ in two $q^2$ intervals $(0.1,1.1)$ and $(1.1,6.0)~{\rm GeV^2}$~\cite{LHCb:2022vje}:
\begin{align}
&{\cal R}_K^{\rm lower}=0.994_{-0.082}^{+0.090}({\rm stat.)}_{-0.027}^{+0.029}({\rm syst.)}, \nonumber \\
&{\cal R}_K^{\rm central}=0.949_{-0.041}^{+0.042}({\rm stat.)}_{-0.022}^{+0.022}({\rm syst.)},
\end{align}
which are compatible with the SM prediction within 0.2 standard deviations and the results superseded previous LHCb measurements. The results reveal that the LU ${\cal R}_K$ is highly consistent both experimentally and theoretically, moreover, this consistency serves as further confirmation of the success and accuracy of the SM in predicting and explaining physical phenomena. However, the branching fraction of $B^+\to K^+\mu^+\mu^-$ still remains somewhat deviant between the data and the SM prediction; thus, it is necessary to continuously provide more precise experimental data and the newly developed SM's predictions to continue refining the SM. In 2016, the BaBar Collaboration searched the FCNC process $B^+\to K^+\tau^+\tau^-$ by using its full data set of $471\times 10^6~ B\bar{B}$ pairs, and reported an upper limit of $\mathcal{B}(B^+\to K^+\tau^+\tau^-)<2.25\times 10^{-3}$ at the $90\%$ confidence level, but no signal was observed~\cite{BaBar:2016wgb}. The SM prediction of ditau channel is approximately $(1\sim  2)\times 10^{-7}$~\cite{Bouchard:2013mia} and the most recent Lattice QCD (LQCD) prediction by the HPQCD Collaboration is expected to be $1.83(13)\times 10^{-7}$~\cite{Parrott:2022zte}. In addition, the BaBar~\cite{BaBar:2008jdv,BaBar:2012mrf}, the CDF~\cite{CDF:2011buy}, the Belle~\cite{Belle:2001oey, Belle:2009zue, BELLE:2019xld}, the CMS~\cite{CMS:2024syx}, and the LHCb~\cite{LHCb:2012juf, LHCb:2014auh, LHCb:2014cxe, LHCb:2016due} Collaborations also conducted researches on the $B^+\to K^+\ell^+\ell^-$ decays for exploring various physical observables, such as the differential branching fractions, the forward-backward asymmetries, the flat term and the isospin asymmetries, etc. Recently, the Belle-II Collaboration reported a measurement of the rare decay $B^+\to K^+\nu\bar{\nu}$ by using the collected 362 ${\rm fb}^{-1}$ samples via the electron-positron collisions~\cite{Belle-II:2023esi}, which has a branching fraction extracted to be $2.3\pm0.5({\rm stat})_{-0.4}^{+0.5}({\rm syst})\times 10^{-5}$, providing the first evidence for this decay at 3.5 standard deviations over the SM prediction from the HPQCD~\cite{Parrott:2022zte}. Compared to the previous measurements by Belle Collaboration~\cite{Belle-II:2021rof, Belle:2017oht}, the new result from Belle-II offers a more precise determination of the boundary for the rare decay $B^+\to K^+\nu\bar{\nu}$.

Theoretically, Refs.~\cite{Huang:2024uuw, Bordone:2024hui, Ali:1999mm,Wu:2006rd, Faessler:2002ut, Bobeth:2011nj, Bobeth:2012vn, Wang:2012ab, Du:2015tda, Li:2018pag, Gubernari:2022hxn, Becirevic:2023aov, Hou:2024vyw, Buras:2021nns, Buras:2022wpw, BhupalDev:2021ipu, Altmannshofer:2009ma, Kamenik:2009kc} offer a multitude of theoretical predictions for the decays $b\to s\ell^+\ell^-$ and $b\to s\nu\bar \nu$ that can afford us access to a deeper horizon. In Ref.~\cite{Huang:2024uuw}, the next-to-leading-order (NLO) weak annihilation contribution is first considered in the rare decays $B\to (K,\pi)\ell^+\ell^-$ with an energetic light flavor meson, which is an important result for SM prediction. Meanwhile, the phenomenological implications of the NLO weak annihilation contribution to the $B\to K \ell^+\ell^+$ decay depend heavily on the light-mesons' leading-twist light-cone distribution amplitudes (LCDAs). Consequently, the precise determination of those leading-twist LCDAs is crucial for investigating those processes. In 2022, the HPQCD Collaboration updated their SM predictions for $B\to K\ell^+\ell^- (\nu\bar{\nu})$ in terms of the LQCD predictions on the $B\to K$ transition form factors (TFFs), which have been calculated across the whole $q^2$-region and have smaller uncertainties than the previous predictions~\cite{Parrott:2022zte}. Moreover, this year, Ref.~\cite{Bordone:2024hui} provides a general amplitude decomposition to improve the performance of testing the SM due to the difficulty of precisely estimating the non-perturbative long-distance contributions related to the charm re-scattering in the $B\to K^{(*)}\ell^+\ell^-$ decays, which describes these effects in full generality in these rare models. Considering the new Belle-II measurement reported with respect to $\mathcal{B}(B^+\to K^+\nu\bar{\nu})$, the authors offer a new model-independent way of introducing lepton flavor-violating coupling and non-diagonal elements of the charged lepton mixing matrix in those models to accommodate the data at the tree level~\cite{Athron:2023hmz}. The resultant Wilson coefficients were mapped into the parameter space of two U($1^\prime$) scenarios, yielding some intriguing results: in the lepton flavor non-violation scenario, the branching fraction $\mathcal{B}(B^+\to K^+\nu\bar{\nu})$ can only generate an increment of about 10 percent; in the lepton flavor violation scenario, on the other hand, once $\mathcal{O}(0.4)$ off-diagonal elements of the charged lepton rotation matrix were included, the $\mathcal{B}(B^+\to K^+\nu\bar{\nu})$ can be large enough.

Since the breaking of factorization caused by photon exchange, the $b\to s\ell^+\ell^-$ transitions are plagued by hadronic uncertainty beyond the TFFs, while the $b\to s\nu\bar \nu$ transition is not affected by it. Moreover, the relation between two processes is not only relevant in the SM, where they are dominated by the same TFFs, but also beyond the SM, due to the SU(2)$_L$ gauge symmetry that relates neutrinos to the left-handed charged leptons~\cite{Buras:2014fpa}. The $B\to K$ TFFs are important hadron elements for studying their physical observables. While the LQCD prediction is, for the most part, reliable, employing different theories to investigate those processes is equally essential for testing and perfecting the SM. It is undeniable that different theoretical methods have distinct applicable $q^2$-regions for TFFs: the perturbative QCD (PQCD) approach is more reliable when the involved energy is hard, $\it i.e.,$ in the large recoil regions; the LQCD results of $B\to$ light meson TFFs are available for soft regions. The QCD light cone sum rules (LCSR) can involve both the hard and the soft contributions below $m_b^2-2m_b\chi$ with $\chi$ refers to a typical hadronic scale of approximately $500 ~{\rm MeV}$, while can be extrapolated to higher $q^2$ regions. As a conclusion, the TFFs results from those three methods are complementary to each other~\cite{Huang:2004hw, Wu:2007vi}. Therefore, we update a scenario on the $B^+\to K^+\ell^+\ell^-(\nu\bar{\nu})$ decays to proceed with a precise calculation for the SM physical observables by applying the QCD LCSR technique.

Meanwhile, the $B\to K$ TFFs are regarded as crucial inputs in the $B\to K\ell^+\ell^-(\nu\bar{\nu})$ FCNC decays, which have been studied within various approaches $\it i.e.$, the relativistic quark model (RQM)~\cite{Faessler:2002ut}, the QCD LCSR~\cite{Lu:2018cfc, Khodjamirian:2010vf, Cui:2022zwm, Ball:2004ye, Wu:2007vi, Wu:2009kq}, the PQCD~\cite{Wu:2007rt, Wang:2012ab}, the AdS/QCD~\cite{Momeni:2017moz} and LQCD~\cite{Bouchard:2013eph, Bailey:2015dka, Parrott:2022rgu}. In which, the QCD LCSR approach allows incorporating information about high-energy asymptotic correlation functions in QCD that parameterize into LCDAs. The meson's LCDAs are universal non-perturbative inputs that enter the exclusive processes involving large momentum transfer $Q^2\gg \Lambda _{{\rm QCD}}^{2}$ and the $B/D$ meson two-body decays through factorization assumptions, while those processes can be decomposed into the long-distance dynamics ($\it i.e.,$ LCDAs) and the perturbatively calculable hard-scattering amplitudes~\cite{Boyle:2006pw, Chetyrkin:2007vm}.  Meanwhile, the contribution of $SU_f(3)$-breaking effect for the $B\to K$ TFFs is about $10\%$, thus a elaborative study on the kaon LCDAs will lead to a better appraisal of these TFFs. In this paper, the kaon leading-twist $\phi_{2;K}(x,\mu_0)$ and twist-3 $\phi_{3;K}^{p,\sigma}(x,\mu_0)$ LCDAs are discussed through coordinating the phenomenological light-cone harmonic oscillator (LHCO) model and the QCD sum rules within the framework of background field theory (BFTSR). By employing this method, the LCDAs for the pseudoscalar mesons $\pi, K, D, \eta^{(\prime)}$~\cite{Zhong:2021epq,Zhong:2022ecl,Zhong:2011rg,Zhong:2022ugk,Hu:2021zmy,Hu:2023pdl}, the scalar mesons $a_0 (980)$, $K_0^* (1430)$~\cite{Wu:2022qqx,Huang:2022xny}, the vector mesons $J/\psi, \rho, \phi$~\cite{Fu:2016yzx,Fu:2018vap,Zhong:2023cyc,Hu:2024tmc}, and the axial meson $a_1(1260)$~\cite{Hu:2021lkl} have been studied in detail.

The remaining portions of the paper are arranged as follows: In Section~\ref{Sec:2}, we introduce the theoretical technique with regard to the FCNC rare decay processes $B\to K\ell^+\ell^-(\nu\bar{\nu})$ for computing the SM observables. Then, the key component TFFs of $B\to K$ decay are calculated based on the QCD LCSR approach, while we introduce briefly the kaon leading-twist $\phi_{2;K}(x,\mu_0)$ and twist-3 $\phi_{3;K}^{p,\sigma}(x,\mu_0)$ LCDAs. Moreover in Section~\ref{Sec:3}, we exhibit our results of $B\to K$ TFFs, while the SM's predictions about the decay $B\to K\ell^+\ell^-(\nu\bar{\nu})$ are also presented with their numerical analysis, including the differential branching fractions, the lepton universality $\mathcal{R}_{K}$ and the angular distribution `flat term' $F_{\rm H}^\ell$. Section~\ref{Sec:4} is a brief summary.

\section{THEORETICAL FRAMEWORK}\label{Sec:2}

Within the SM, the rare $B$ decays can be dealt with by using the effective Hamiltonian, which contains operators $\mathcal{O}_i$ that contribute to the $b\to q\ell\ell$ and $b\to q\gamma$ with $q=(s, d)$ transitions at the quark level. The effective Hamiltonian can be written as~\cite{Blake:2016olu}:
\begin{align}
\mathcal{H} _{{\rm eff}}^{b\rightarrow q}&=\frac{4G_F}{\sqrt{2}} \bigg\{ \sum_{i=1}^2{[ \lambda _{\mu}^{(q)}C_i\mathcal{O}_{i}^{u}+\lambda _{c}^{(q)}C_i\mathcal{O}_{i}^{c} ]}-\lambda _{t}^{(q)}\sum_{i=3}^{10}{C_i\mathcal{O}_i} +{\rm h.c.} \bigg\},
\end{align}
where $G_F=1.1663787(6)\times 10^{-5}$ refers to the Fermi coupling constant, $\lambda_{p}^{q}=V_{pb}V_{pq}^*$ stands for the CKM matrix element, and the Wilson coefficients can be computed perturbatively, whose values shall be alternated when new particles beyond the SM are included. The operator basis have the following definitions~\cite{Bobeth:1999mk, Altmannshofer:2008dz}:
\begin{eqnarray}
&\mathcal{O} _{1}^{p}&=(\bar{q}_L\gamma _{\mu}T^ap_L) (\bar{p}_L\gamma ^{\mu}T^ab_L),
\\
&\mathcal{O} _{2}^{p}&=(\bar{q}_L\gamma _{\mu}p_L) (\bar{p}_L\gamma ^{\mu}b_L),
\\
&\mathcal{O} _3&=(\bar{q}_L\gamma _{\mu}b_L) \sum_p{(\bar{p}\gamma ^{\mu}p)},
\\
&\mathcal{O} _4&=(\bar{q}_L\gamma _{\mu}T^ab_L) \sum_p{(\bar{p}\gamma ^{\mu}T^ap)},
\\
&\mathcal{O} _5&=(\bar{q}_L\gamma _{\mu}\gamma _{\nu}\gamma _{\rho}b_L) \sum_p{( \bar{p}\gamma ^{\mu}\gamma ^{\nu}\gamma ^{\rho}p)},
\\
&\mathcal{O} _6&=(\bar{q}_L\gamma _{\mu}\gamma _{\nu}\gamma _{\rho}T^ab_L) \sum_p{( \bar{p}\gamma ^{\mu}\gamma ^{\nu}\gamma ^{\rho}T^ap)},
\\
&\mathcal{O} _7&=\frac{e}{16\pi ^2}m_b(\bar{q}_L\sigma ^{\mu \nu}b_R) F_{\mu \nu},
\\
&\mathcal{O} _8&=\frac{g_s}{16\pi ^2}m_b(\bar{q}_L\sigma ^{\mu \nu}T^ab_R) G_{\mu \nu}^{a},
\\
&\mathcal{O} _9&=\frac{e^2}{16\pi ^2}(\bar{q}_L\gamma _{\mu}b_L) \sum_{\ell}{( \bar{\ell}\gamma ^{\mu}\ell)},
\\
&\mathcal{O} _{10}&=\frac{e^2}{16\pi ^2}(\bar{q}_L\gamma _{\mu}b_L) \sum_{\ell}{( \bar{\ell}\gamma_5\ell)},
\end{eqnarray}
where $L$ and $R$ are left-handed and right-handed chiralities of the fermions with $q=(u, c)$, and $\ell=(e, \mu, \tau)$, while the sum runs over all the five active quark flavors, {\it i.e.} $p=(u,d,s,c,b)$. For the values of the Wilson coefficients at the $b$-quark scale $\mu_b\sim {\cal O}(m_b)$, we adopt the ones suggested in Ref.~\cite{Huber:2005ig}, and for convenience we put their values together with the uncertainties in Table~\ref{Tab:Wilson Coefficients}.

\begin{table*}[t]
\footnotesize
\begin{center}
\caption{The adopted input values for the Wilson coefficients at the $b$-quark scale $\mu_b$~\cite{Huber:2005ig}.}
\label{Tab:Wilson Coefficients}
\begin{tabular}{lllll}
\hline
$C_1$~~~~~~~~~~~~~~~~~~&$C_2$~~~~~~~~~~~~~~~~~~&$C_3$~~~~~~~~~~~~~~~~~~&$C_4$~~~~~~~~~~~~~~~~~~&$C_5$\\
$-0.29(16)$ &$1.009(10)$ &$-0.0047(42)$ &$-0.081(39)$ &$0.00036(31)$ \\
\hline
$C_6$&$C_7$&$C_8$&$C_9$&$C_{10}$
\\
$0.00082(97)$ &$-0.297(26)$ &$-0.152(15)$ &$4.04(33)$ &$-4.292(73)$ \\
\hline
\end{tabular}
\end{center}
\end{table*}

The SM differential decay rate of $B\to K\ell^+\ell^-$ with $\ell= (e,\mu,\tau)$ is constructed, which has following form~\cite{Becirevic:2012fy}:
\begin{eqnarray}\label{Eq:dGamma}
\frac{d\Gamma _{\ell}}{dq^2}=2a_{\ell}(q^2)+\frac{2}{3}c_{\ell}(q^2),
\end{eqnarray}
with the indices $a_{\ell}(q^2)$ and $c_{\ell}(q^2)$ can be expressed as
\begin{align}\nonumber
a_{\ell}(q^2)&=\mathcal{C} \bigg\{ q^2|F_P(q^2)|^2+\frac{\lambda (q,m_B,m_K)}{4} \Big[|F_A(q^2)|^2+|F_V(q^2)|^2 \Big]
\\
&+4m_{\ell}^{2}m_B^2|F_A(q^2)|^2+2m_{\ell} (m_B^2-m_{K}^{2}+q^2) {\rm Re}(F_P(q^2) F_{A}^{*}(q^2)) \bigg\},
\\
c_{\ell}(q^2)&=-\frac{\mathcal{C} \lambda (q^2,m_B,m_K) \beta _{\ell}^{2}}{4}\Big[ |F_A(q^2)|^2+|F_{V}^{2}(q^2)|\Big],
\end{align}
where
\begin{align}
& \mathcal{C} =\frac{G_F^2 \alpha_{\rm ew}^2 |V_{tb}V_{ts}^*|^2}{2^9\pi^5 m_B^{3}}\beta_\ell \sqrt{\lambda(q,m_B,m_K)},
\\
& \beta _{\ell}=\sqrt{1-\frac{4m_{\ell}^{2}}{q^2}},
\\
& \lambda (a,b,c) =a^4+b^4+c^4-2(a^2b^2+a^2c^2+b^2c^2).
\end{align}
The helicity form factors $F_{P,V,A}(q^2)$ can be represented by the scalar, vector and tensor TFFs $f_{+,0,\rm T}^{BK}(q^2)$, which are
\begin{align}
&F_P(q^2)=-m_{\ell}C_{10}(\mu_b) \left[ f_+^{BK}(q^2) -\frac{m_B^2-m_{K}^{2}}{q^2}( f_0^{BK}(q^2) -f_+^{BK}(q^2)) \right],
\\
&F_V(q^2)=C_{9}^{{\rm eff}}(\mu_b) f_+^{BK}(q^2) +\frac{2m_b}{m_B+m_K}C_{7}^{{\rm eff}}(\mu_b) f_{\rm T}^{BK}(q^2),
\\
&F_A(q^2)=C_{10}f_+^{BK}(q^2).
\end{align}
The above formulas together with the explanations of notations can be found in Refs.~\cite{Parrott:2022zte, Becirevic:2012fy}. The Wilson coefficients $C_7^\mathrm{eff}$ and $C_9^\mathrm{eff}$ are important inputs, and they have relatively complex expressions when considering more precise calculations. We do not present the formulas of them here, which can be read from Refs.~\cite{Bobeth:2011nj, Bobeth:2012vn, Beylich:2011aq, Beneke:2001at, Greub:2008cy, Grinstein:2004vb, Bobeth:2010wg, Asatryan:2001zw, Asatrian:2001de}. The interested readers may turn to those references for more detail. Our intention here is to assess the significance of those corrections to the physical observables. For the other input parameters, we adopt $\alpha_{\rm ew}\approx 1/137$, the CKM matrix element $|V_{tb}V_{ts}^{*}|=0.04185(93)$~\cite{Dowdall:2019bea}, and the lifetime of initial state $B^+$-meson $\tau_{B^+}=1.638(4)$ ps~\cite{HFLAV:2022esi}.

In the $b\to s\ell^+\ell^-$ decays, there are many fascinating phenomena, one of which is LU. Examining the LU breaking in $b\to s\ell^+\ell^-$ decays would provide an unambiguous indication of new physics beyond the SM~\cite{LHCb:2022vje}. Therefore, the relative decay rates to muon and electron final states, integrated over a region of the square of the dilepton invariant mass $q_a^2\leqslant q^2\leqslant q_b^2$, are used to construct the ratio ${\cal R}_K$, i.e.
\begin{align}\label{Eq:RK}
{\cal R}_K(q_a^2,q_b^2) =\int_{q_a^2}^{q_b^2}\dfrac{{\rm d}\Gamma (B^+\to K^+\mu ^+\mu ^-)}{{\rm d}q^2}{\rm d}q^2 \bigg/
\int_{q_a^2}^{q_b^2}{\dfrac{{\rm d}\Gamma (B^+\to K^+e^+e^-)}{{\rm d}q^2}{\rm d}q^2}.
\end{align}
Regarding the LU, we will calculate it in two $q^2$-intervals, i.e., the lower $q^2$ range: $(0.1 \leqslant q^2\leqslant 1.1)~{\rm GeV^2}$ and the central $q^2$ range: $(1.1\leqslant q^2\leqslant 6.0)~{\rm GeV^2}$.

Moreover, the angular distribution is also an important observable for testing the existence of possible signal beyond the SM. Two additional experimentally measured quantities with which we can compare are based on studying the angular distribution of decay products. The differential distribution in angle can be written as~\cite{Bobeth:2007dw}:
\begin{eqnarray}
\frac{1}{\Gamma _{\ell}}\frac{d\Gamma _{\ell}}{d\cos \theta}=\frac{3}{4}( 1-F_{\rm H}^\ell) (1-\cos ^2\theta) +\frac{1}{2}F_{\rm H}^\ell+A_{{\rm FB}}^\ell\cos \theta,
\end{eqnarray}
where $\theta$ is the angle between $B$ and $\ell$ as measured in the dilepton rest frame. Here we have a notation that both the flat term $F_{\rm H}^\ell/2$ and the forward-backward asymmetry $A_{{\rm FB}}^\ell$ give small contributions, which are sensitive to new physics. Due to $A_{{\rm FB}}^\ell$ is equals zero in SM, so we will not consider it here. According to Ref.~\cite{Bobeth:2007dw}, we have
\begin{eqnarray}
F_{\rm H}^\ell(q_{\rm lower}^{2},q_{\rm up}^2) =\int_{q_{{\rm lower}}^{2}}^{q_{{\rm up}}^{2}}{(a_{\ell}+c_{\ell}) dq^2} \bigg/ \int_{q_{{\rm lower}}^{2}}^{q_{{\rm up}}^{2}}{(a_{\ell}+c_{\ell}/3) dq^2}
\end{eqnarray}

Lastly, we make a study on the rare decay $B^+\to K^+\nu\bar{\nu}$, which is currently of particular interest. Accurate observations in this channel may provide fresh insights into the mechanics behind the violation of lepton-flavor universality. The FCNC rare decays $b\to s\nu\bar{\nu}$ are described by the effective Hamiltonian~\cite{Blake:2016olu}:
\begin{eqnarray}
\mathcal{H} _{{\rm eff}}^{b\to s\nu\bar{\nu}} =\frac{4G_F}{\sqrt{2}}\lambda _t\sum_a{C_a\mathcal{O} _a}+{\rm h}.{\rm c}.,
\end{eqnarray}
where the only relevant operator in the SM is
\begin{eqnarray}
\mathcal{O} _{L}^{\nu _i\nu _j}=\frac{e^2}{(4\pi) ^2}(\bar s_L\gamma _{\mu}b_L) ( \bar{\nu}_i\gamma ^{\mu}(1-\gamma_5) \nu _j) .
\end{eqnarray}
The effective Wilson coefficient in SM is $[ C_{L}^{\nu _i\nu _j} ] _{{\rm SM}}\equiv \delta _{ij}C_{L}^{{\rm SM}}$ with $C_{L}^{{\rm SM}}=-X_t/\sin ^2\theta _W$. Here, $X_t=1.468(17)$ refers to the top-quark contribution, which is computed the full electroweak two-loop correction~\cite{Brod:2010hi} and $\sin ^2\theta _W=0.2314(4)$ is the Weinberg angle~\cite{ParticleDataGroup:2022pth}. Therefore, one can obtain $C_{{\rm L}}^{{\rm SM}}=6.34(7) $ and its the dominant source of uncertainty from the higher order QCD corrections. The SM differential decay rate of $B^+\to K^+\nu\bar{\nu}$ can be expressed as~\cite{Becirevic:2023aov}:
\begin{align}
&\frac{{\rm d}\mathcal{B}}{{\rm d}q^2}(B\rightarrow K\nu \bar{\nu})=\mathcal{N} _K(q^2) \left| C_{{\rm L}}^{{\rm SM}} \right|^2|\lambda _t|^2\left| f_+^{BK}(q^2) \right|^2,
\end{align}
where $\mathcal{N}_K$ is a $q^2$-dependent function, which denotes as:
\begin{eqnarray}
\mathcal{N} _K(q^2) =\tau _{B^+}\frac{G_{F}^{2}\alpha _{{\rm ew}}^{2}\lambda ( q,m_B,m_K) ^{3/2}}{256\pi ^5m_{B}^{3}},
\end{eqnarray}
and $\lambda_{t}=|V_{tb}V_{ts}^{*}|$ is the CKM element. The remaining input parameters can be found in the preceding part. Given the disparity that exists between the SM predictions and the recent experimental results from the Belle-II Collaboration, it is required to provide the SM observations regarding the rare decay $B^+\to K^+\nu\bar{\nu}$.

The $B\to K$ TFFs $f_+^{BK}(q^2)$, $f_0^{BK}(q^2)$ and $f_{\rm T}^{BK}(q^2)$ are important inputs for the above formulas. To obtain their analytic expressions under the LCSR approach, we consider the vacuum-to-kaon correlation function as follows:
\begin{eqnarray}
\Pi _{\mu}(p,q) =i\int{d^4x}e^{iq\cdot x}\langle K(p) |T \{ j_{\mu}(x) ,j_{5}^{\dagger}(x)\} |0\rangle ,
\end{eqnarray}
where the current $j_{\mu}(x) =\bar s(x) \Gamma _{\mu}b(x)$ with $\Gamma _{\mu}=\gamma _{\mu}$ or $\Gamma _{\mu}=-i\sigma_{\mu\nu}q^{\nu}$ corresponds to two types of $b\to s$ transitions, respectively, while $j_{5}^{\dagger}(0) =m_b\bar{b}(0) i\gamma_5u(0)$. For the large virtualities of the currents above, the correlation function is dominated by the distances near the light cone $x^2\thickapprox0$, which can be factorized into the convolution of the non-perturbative but universal part with the perturbatively calculable, short-distance part. One can relate the correlation function to the $B\to K$ matrix elements based on the hadronic dispersion relation in the virtuality $(p+q)^2$, which have the following forms:
\begin{align}
\langle K(p) &|\bar s\gamma _{\mu}b|B(p+q) \rangle =2f_+^{BK}(q^2) p_{\mu}+[ f_+^{BK}(q^2) +f_-^{BK}(q^2) ] q_{\mu},
\label{Eq:28} \\
\langle K(p) &|\bar s\sigma_{\mu\nu}q^{\nu}b|B(p+q) \rangle =[ q^2( 2p_{\mu}+q_{\mu}) -(m_B^2-m_{K}^{2}) q_{\mu} ] \frac{if_{\rm T}^{BK}(q^2)}{m_B+m_K},
\label{Eq:29}
\end{align}
where the TFFs can be written as a series over the $K$-meson LCDAs with increasing twists. On the other hand, in the time-like region, by inserting the complete intermediate state with the same quantum numbers as the current operator into the hadron current of the correlation function and isolating the pole term of the $B$-meson ground-state contributions for all three invariant amplitudes $F[q^2,(p+q)^2]$, $\tilde{F}[q^2,(p+q)^2]$, and $F^{\rm T}[q^2,(p+q)^2]$ while using Eqs.~(\ref{Eq:28}) and (\ref{Eq:29}), we obtain
\begin{align}
f_+^{BK}(q^2) &=\frac{e^{m_B^2/M^2}}{2m_B^2f_B}
\left[ F_0(q^2,M^2,s_0) +\frac{\alpha _sC_F}{4\pi}F_1(q^2,M^2,s_0) \right],
\label{Eq:fp}
\\
f_+^{BK}(q^2) &+f_-^{BK}(q^2) =\frac{e^{m_B^2/M^2}}{m_B^2f_B}\left[ \tilde{F}_0(q^2,M^2,s_0) +\frac{\alpha _sC_F}{4\pi}\tilde{F}_1(q^2,M^2,s_0) \right],
\label{Eq:fpm}
\\
f_{\rm T}^{BK}(q^2) &=\frac{(m_B+m_K) e^{m_B^2/M^2}}{2m_B^2f_B}\left[ F_{0}^{\rm T}(q^2,M^2,s_0) +\frac{\alpha _sC_F}{4\pi}F_{1}^{\rm T}(q^2,M^2,s_0) \right],
\label{Eq:fT}
\end{align}
where the usual Borel transformation have been implicitly done to suppress the higher-twist contributions and the less certain continuum contributions. Moreover, the scalar $B\to K$ TFF is a combination of $f_{+}^{BK}(q^2)$ and $f_{-}^{BK}(q^2)$:
\begin{eqnarray}
f_0^{BK}(q^2) =f_+^{BK}(q^2) +\frac{q^2}{m_B^2-m_{K}^{2}}f_-^{BK}(q^2).
\end{eqnarray}
In above formulas, the invariant amplitudes $F_{0(1)}, \tilde{F}_{0(1)}$, and $F_{0(1)}^{\rm T}$ refer to the LO (NLO) contributions, respectively. Those functions have similar expressions as those of the LO TFFs~\cite{Duplancic:2008tk} and the NLO TFFs~\cite{Duplancic:2008ix}. We have verified by cross-check that using the appropriate transformations, we can obtain the same results as those references. Therefore, we do not offer the complex expressions of those invariant amplitudes here in order to shorten the manuscript's length.

The dominant calculation of $B\to K$ TFFs are primarily influenced by the kaon leading-twist $\phi_{2;K}(x,\mu_0)$ and twist-3 $\phi_{3;K}^{p,\sigma}(x,\mu_0)$ LCDAs, where $\mu_0$ represents some initial scale which is usually taken as $1$ GeV. For the kaon leading-twist and twist-3 LCDAs, they can be constructed by the phenomenological LHCO model and the framework of BFTSR, which have been studied in Refs.~\cite{Zhong:2021epq, Zhong:2022ecl}. They are connected with the kaon leading-twist and twist-3 wavefunctions (WFs) as
\begin{eqnarray}
\phi_{i;K}(x,\mu)={\cal N}_i\int_{|\mathbf{k}_\bot|\leqslant \mu^2}\frac{d^2\mathbf{k}_\bot}{16\pi^3}\Psi_{i;K}(x,\mathbf{k}_\bot),
\end{eqnarray}
with $\mathbf{k}_\bot$ is the kaon transverse momentum and the index $i=(2,3)$ stands for the leading-twist and twist-3 LCDAs, respectively. The normalization coefficients refer to ${\cal N}_2=2\sqrt{6}/f_K$ and ${\cal N}_3=1$, while the functions $\Psi_{i;K}(x,\mathbf{k}_\bot)$ with $i=(2,3)$ are the WFs for the kaon leading-twist and twist-3, respectively. They have been constructed using the LCHO model, and the detailed procedures have been given in Refs.~\cite{Zhong:2021epq, Zhong:2022ecl}. Here, we only present the final formulas for the leading-twist $\phi_{2;K}(x,\mu)$ and twist-3 $\phi_{3;K}^{p,\sigma}(x,\mu)$ LCDAs. The leading-twist LCDA $\phi_{2;K}(x,\mu)$ takes the form:
\begin{eqnarray}
\phi _{2;K}(x,\mu) &=& \frac{\sqrt{3}A_{2;K}\beta _{2;K}\tilde{m}}{2\pi ^{3/2}f_K}\sqrt{x\bar x}\varphi _{2;K}(x)\exp \left[ -\frac{\hat m_q^2x+\hat{m}_{s}^{2}\bar x -\tilde{m}^2}{8\beta _{2;K}^{2}x\bar x} \right] \nonumber \\
& &\times\left\{ {\rm Erf}\left[ \sqrt{\frac{\tilde{m}^2+\mu ^2}{8\beta _{2;K}^{2}x\bar x}} \right] -{\rm Erf}\left[ \sqrt{\frac{\tilde{m}^2}{8\beta _{2;K}^{2}x\bar x}} \right] \right\}, \label{T2}
\end{eqnarray}
where $\bar{x}=1-x$, $\tilde{m}=\hat{m}_qx+\hat{m}_s\bar{x}$, and $\varphi_{2;K}(x)$ dominates the behavior of the WF's longitudinal distribution, which can be expressed as $\varphi_{2;K}(x)=(x\bar{x})^{\alpha_{2;K}}[1+0.4\hat{B}_2^{2;K}C_1^{3/2}(\xi) +\hat{B}_2^{2;K}C_2^{3/2}(\xi)]$ with $C_{n}^{3/2}(\xi)$ being the Gegenbauer polynomial, where $\xi = (2x-1)$. The twist-3 LCDA $\phi_{3;K}^{p,\sigma}(x,\mu)$ takes the form:
\begin{align}\label{T3}
\phi_{3;K}^{p,\sigma}(x,\mu) &= \frac{A_{3;K}^{p,\sigma} (\beta_{3;K}^{p,\sigma})^2}{2\pi^2} \varphi_{3;K}^{p,\sigma}(x)
\exp \left[ -\frac{1}{8(\beta_{3;K}^{p,\sigma})^2} \left(\frac{\hat{m}_q^2}{x}+\frac{\hat{m}_s^2}{\bar x}\right)\right]
\left[ 1-\exp \left( -\frac{\mu^{2}}{8(\beta_{3;K}^{p,\sigma})^2x\bar x} \right) \right],
\end{align}
with $\varphi _{3;K}^{\left( p,\sigma \right)}\left( x \right) =1+\hat{B}_{3;K}^{\left( p,\sigma \right)}C_{1}^{1/2}\left( \xi \right) +\hat{C}_{3;K}^{\left( p,\sigma \right)}C_{2}^{3/2}\left( \xi \right)$. Here, the selection of the constituent quark masses $\hat{m}_q$ and $\hat{m}_s$ is different in Eqs.~(\ref{T2}) and (\ref{T3}), whose values are discussed specifically in Refs.~\cite{Zhong:2021epq,Zhong:2022ecl}. Moreover, while the contribution of TFFs from the kaon twist-4 LCDAs is somewhat small, they are an indispensable part of accurate calculation. Therefore, we also consider their contributions to the TFFs in this manuscript, whose formulas can be found in Ref.~\cite{Ball:2006wn}.

\section{Numerical Analysis}\label{Sec:3}

In processing the numerical analysis, we take the following basic input parameters from Particle Data Group (PDG)~\cite{ParticleDataGroup:2022pth}: $m_B=5279.45\pm0.08$ MeV, $m_K=493.677\pm0.013$ MeV, the $u$ and $s$-quark masses at scale $\mu=2$ GeV are adopted as $m_u=2.16_{-0.26}^{+0.49}$ MeV and $m_s=93_{-5}^{+11}$ MeV, respectively, while the $b$-quark mass $m_b(\mu_b)=4.18_{-0.02}^{+0.03}$ GeV. The $K$-meson decay constant $f_K/f_\pi=1.1932$~\cite{FlavourLatticeAveragingGroup:2019iem} with $f_\pi=130.2(1.2)$ MeV~\cite{ParticleDataGroup:2022pth}. The $B$-meson decay constant $f_B=214\pm18~\mathrm{MeV}$ was computed in the $\overline{\rm MS}$ scheme by using the QCD sum rule expressions at the $\mathcal O(\alpha_s,m_s^2)$-order level~\cite{Duplancic:2008tk,Jamin:2001fw}.

\begin{table}[t]
\footnotesize
\begin{center}
\caption{The several typical model parameters of kaon leading-twist and twist-3 LCDAs at scale $\mu_0 =1$~GeV and $\mu_k = 3$~GeV.}
\label{Tab:ModelParameters}
\begin{tabular}{l l l l l }
\hline
~~~~~~~~~~~~~~& $A_{2;K}(\mathrm{GeV^{-1}})$~~~~~& $\alpha_{2;K}$~~~~~~~& $\hat{B}_{2}^{2;K}$~~~~~~~&$\beta_{2;K}(\mathrm{GeV})$\\
$\mu_0$             &4.088       &$-1.068$      &$-0.113$      &0.681           \\
$\mu_k$             &10.20       &0.003       &0.008       &1.117            \\
\hline
~~~~&$A_{3;K}^{p}(\mathrm{GeV^{-1}})$~~~~~&$B_{3;K}^{p}$~~~~~~~&$C_{3;K}^{p}$~~~~~~~&$\beta_{3;K}^{p}(\mathrm{GeV})$\\
$\mu_0$             &143.5       &$-0.019$         &1.583          &0.546        \\
$\mu_k$             &138.4       &0.037          &1.521          &0.538         \\
\hline
~~~~&$A_{3;K}^{\sigma}(\mathrm{GeV^{-1}})$~~~~~&$B_{3;K}^{\sigma}$~~~~~~~&$C_{3;K}^{\sigma}$~~~~~~~&$\beta_{3;K}^{\sigma}(\mathrm{GeV})$\\
$\mu_0$             &154.7       &0.010          &0.237          &0.477        \\
$\mu_k$             &172.5       &0.027          &0.181          &0.455        \\
\hline
\end{tabular}
\end{center}
\end{table}

\begin{table}[t]
\footnotesize
\begin{center}
\caption{TFFs $f_{+;0;\rm T}^{BK}(0)$ at the large recoil point, containing error sources in quadrature.}
\label{Tab:TFFsEndPoint}
\begin{tabular}{l l l l l}
\hline
$\rm{Sources}$~~~~~~~~& $f_{+}^{BK}(0)$~~~~~~~~~~~~~& $f_{0}^{BK}(0)$~~~~~~~~~~~~~& $f_{\rm T}^{BK}(0)$\\
\hline
$s_0$               &$_{-0.010}^{+0.010}$             &$_{-0.010}^{+0.010}$          &$_{-0.011}^{+0.010}$   \\
$M^2$               &$_{-0.004}^{+0.004}$             &$_{-0.004}^{+0.004}$          &$_{-0.002}^{+0.002}$    \\
$f_B$               &$_{-0.025}^{+0.030}$             &$_{-0.025}^{+0.030}$          &$_{-0.021}^{+0.025}$     \\
$m_b$               &$_{-0.002}^{+0.002}$             &$_{-0.002}^{+0.002}$          &$_{-0.002}^{+0.002}$      \\
$\mathrm{Total}$    &$0.328_{-0.028}^{+0.032}$        &$0.328_{-0.028}^{+0.032}$      &$0.277_{-0.024}^{+0.028}$  \\
\hline
\end{tabular}
\end{center}
\end{table}

The model parameters of the leading-twist and twist-3 LCDAs at the scale $\mu_0=1~{\rm GeV}$ have been determined in our previous works, but to correspond to the applicable energy scale $( m_{B}^{2}-m_{b}^{2})^{1/2}\approx 3~{\rm GeV}$ for $B\to K$ TFFs, we recalculate the values for these model parameters at the scale $\mu_k=3~{\rm GeV}$ in Table~\ref{Tab:ModelParameters}. With the resultant determined twist-2 and twist-3 LCDAs, one can further proceed the $B\to K$ TFFs. To get the $B\to K$ TFFs QCD sum rule parameters, we take the following four criteria: (1) The continuum threshold $s_0^B$, serving as the demarcation between the $B$-meson ground state and higher mass contributions, is usually set to the value that is close to the first known resonance of the $B$-meson ground state. Therefore, set $s_0^B$ to $34.0\pm 1.0$ GeV$^2$, which indicates that the excitation energy ranges from approximately 0.45 GeV to 0.65 GeV. (2) When we expand the correlator over $1/M^2$ and calculate it as an all-power series, it should be independent of the choice of $1/M^2$. However, since we only know its first several terms, we have to set a proper range for $M^2$. (3) As a conservative prediction, we require the continuum contribution to be less than 30\% of the total LCSR in order to set the upper limit of $M^2$. (4) Furthermore, we also require that the contribution from twist-4 LCDAs does not exceed 5\%. At these points, we can properly determine the two parameters. Then, we get the suitable Borel parameters $M^2$ and continuum threshold $s_0^B$ for $B\to K$ TFFs $M^2=22\pm 1$ GeV$^2$ and $s_0^B=34 \pm 1$ GeV$^2$, respectively.

\begin{table}[t]
\footnotesize
\begin{center}
\caption{The fitted parameters with TFFs $f_{+;0;\rm T}^{BK}(q^2)$ from the SSE method, including the central, upper and lower limits.}
\label{Tab:SSEcoefficients}
\begin{tabular}{llll}
\hline
$f_{+}^{BK}(q^2)$~~~~~~~~~~~~~~~~~~~~~&Central~~~~~~~~~~~~~~~~~~~~~&Upper~~~~~~~~~~~~~~~~~~~~~&Lower \\
\hline
$\beta_{1}^+$    & $+0.325$           & $+0.357$          & $+0.297$      \\
$\beta_{2}^+$    & $-0.966$           & $-1.055$          & $-0.906$       \\
$\beta_{3}^+$    & $-1.533$           & $-1.750$          & $-1.460$        \\
$\Delta$       & $0.415\%$          & $0.432\%$         & $0.407\%$        \\
\hline
$f_{0}^{BK}(q^2)$&Central&Upper&Lower \\
\hline
$\beta_{1}^0$    & $+0.329$           & $+0.361$         & $+0.300$      \\
$\beta_{2}^0$    & $+0.329$           & $+0.369$         & $+0.284$        \\
$\beta_{3}^0$    & $-0.104$           & $-1.150$         & $-0.995$        \\
$\Delta$       & $0.060\%$          & $0.070\%$        & $0.055\%$        \\
\hline
$f_{\rm T}^{BK}(q^2)$&Central&Upper&Lower  \\
\hline
$\beta_{1}^{\rm T}$    & $+0.270$           & $+0.298$         & $+0.247$      \\
$\beta_{2}^{\rm T}$    & $-1.472$           & $-1.624$         & $-1.355$       \\
$\beta_{3}^{\rm T}$    & $-2.742$           & $-3.182$         & $-2.50$        \\
$\Delta$       & $0.817\%$          & $0.842\%$        & $0.811\%$        \\
\hline
\end{tabular}
\end{center}
\end{table}

Furthermore, we calculate the $B\to K$ TFFs at large recoil point using the determined parameters, which are exhibited in Table~\ref{Tab:TFFsEndPoint}, where the errors of all the mentioned error sources have been added up in quadrature. Theoretically, the LCSR approach can effectively apply in lower and intermediate $q^2$-region. Thus, we go by the physical allowable range $0\leqslant q^2 \leqslant (m_B-m_K)^2$ of the $B \to K$ transition to conclude a corresponding region, $\it{i.e.}$ $q^2\in [0,10.0]$ GeV$^2$. In order to observe the behavior of the TFFs across the entire physically allowable $q^2$-region, one can utilize the simplified series expansion (SSE) method, which satisfies the following formulae:
\begin{eqnarray}
f_{+;0;\rm T}^{BK}(q^2) =\frac{1}{1-q^2/m_{B}^{2}}\sum_{k=1,2,3}{\beta_k^{+;0;\rm T}} z^k (q^2,t_0),
\end{eqnarray}
in which $\beta_k^{+;0;\rm T}$ are real coefficients and function $z^k(q^2,t_0)=(\sqrt{t_+ -q^2}-\sqrt{t_+ -t_0})/(\sqrt{t_+ -q^2}+\sqrt{t_+ -t_0})$ with $t_\pm =(m_B\pm m_K)^2$ and $t_0=t_\pm (1-\sqrt{1-t_-/t_+})$. With the help of the SSE method, one can keep the analytic structure correct in the complex plane and ensure the appropriate scaling, $f_{+;0;\rm T}^{BK}(q^2)\sim  1/q^2$ at large $q^2$-region. Meanwhile, we use the quality of fit $\Delta$ to take stock of the resultant of extrapolation, which is defined as:
\begin{eqnarray}
\Delta =\frac{\sum_t{|F_i\left( t \right) -F_{i}^{\mathrm{fit}}\left( t \right) |}}{\sum_t{|F_i\left( t \right) |}}\times 100
\end{eqnarray}
The reasonability request for the $\Delta$ is less than one percent based on the fitted parameters $\beta_k^{+;0;\rm T}$ with indices $k=1,2,3$ and we exhibit them in Table~\ref{Tab:SSEcoefficients}.

\begin{figure}[t]
\begin{center}
\includegraphics[width=0.495\textwidth]{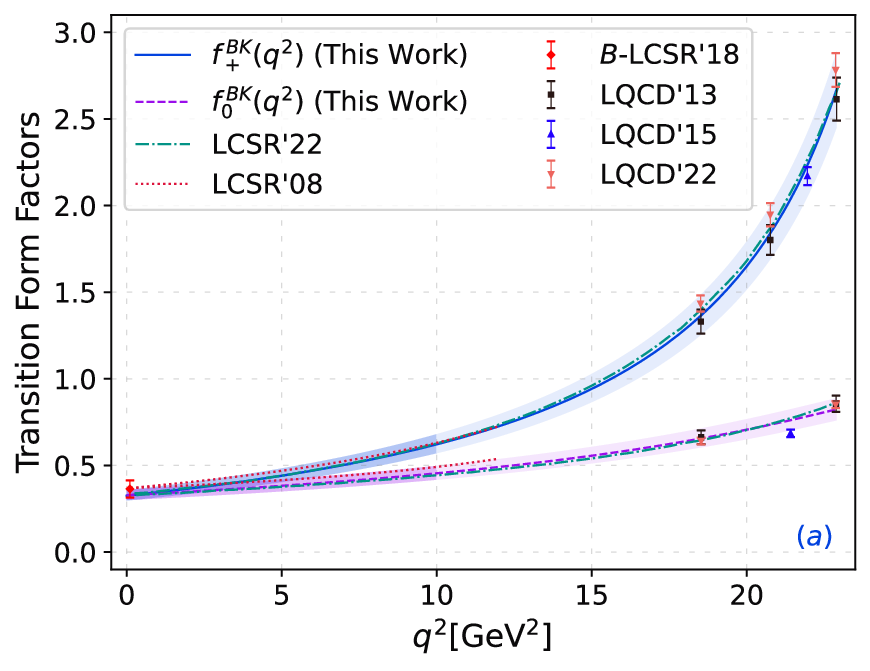}
\includegraphics[width=0.495\textwidth]{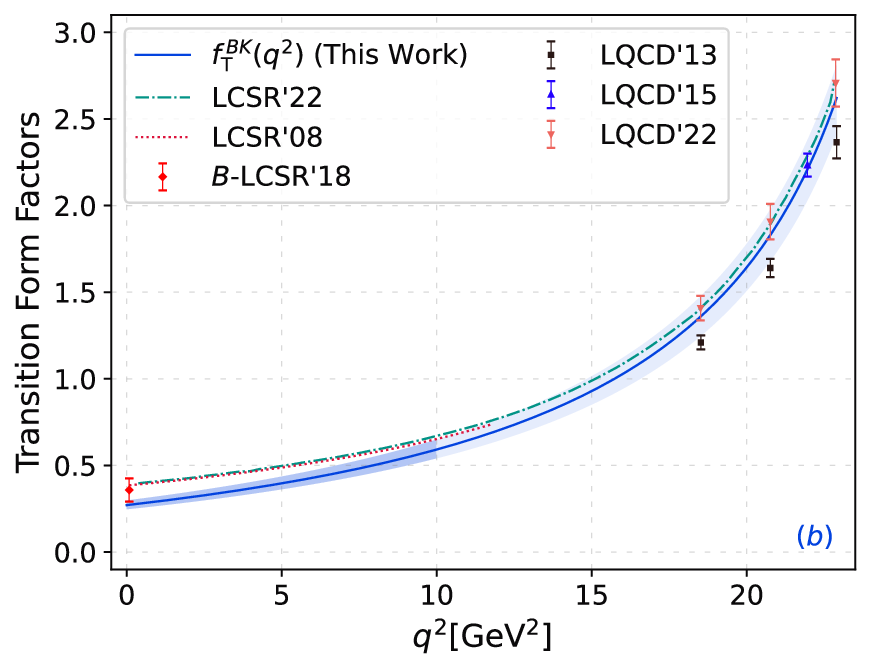}
\end{center}
\caption{Plotting of $B\to K$ TFFs $f_{+,0,\rm T}^{BK}(q^2)$, along with associated uncertainties, is done throughout the whole $q^2$-region. In the meantime, we also displayed other results in various $q^2$ interval for comparison $\it i.e.,$ the LCSR'22 by Cui~\cite{Cui:2022zwm}, the LCSR'08 by Duplancic~\cite{Duplancic:2008tk}, BLCSR'18 by Lu~\cite{Lu:2018cfc}, the LQCD'13, LQCD'15, and LQCD'22 from Lattice QCD~\cite{Bouchard:2013eph}, Three-Flavor Lattice QCD~\cite{Bailey:2015dka}, Relativistic Lattice QCD~\cite{Parrott:2022rgu}, respectively.}
\label{Fig:TFFs_Fp_Fz}
\end{figure}

The behaviors of the extrapolated TFFs $f_{+}^{BK}(q^2)$, $f_{0}^{BK}(q^2)$, and $f_{\rm T}^{BK}(q^2)$ in whole physical $q^2$-region are shown in Fig.~\ref{Fig:TFFs_Fp_Fz}. Where the shaded bands manifest uncertainties from different input parameters, and the darker band refers to the result of the LCSR prediction, while the lighter band is the SSE prediction. Meanwhile, we also present other results in various $q^2$ interval for comparison. Additionally, we make the following statements:
\begin{itemize}
    \item   The behaviors of TFFs $f_{+}^{BK}(q^2)$ and $f_{0}^{BK}(q^2)$ in Fig.~\ref{Fig:TFFs_Fp_Fz} show a good agreement with the LCSR'22 by Cui~\cite{Cui:2022zwm}, the LCSR'08 by Duplancic~\cite{Duplancic:2008tk}, the LQCD'13, LQCD'15, and LQCD'22 results from the Lattice QCD~\cite{Bouchard:2013eph}, Three-Flavor Lattice QCD~\cite{Bailey:2015dka}, and Relativistic Lattice QCD~\cite{Parrott:2022rgu}, respectively. Moreover, the results from the $B$-LCSR'18 ($B$-meson LCSR) by Lu~\cite{Lu:2018cfc}, slightly deviate from our predictions at large recoil point. For TFF $f_{0}^{BK}(q^2)$, we just chose two points toward the tail end for comparison because the uncertainty of the LQCD result is small and they're in our shadow band.
    \item   The behavior of TFF $f_{\rm T}^{BK}(q^2)$ in Fig.~\ref{Fig:TFFs_Fp_Fz} (b) has a well accord with the LCSR'22, LCSR'08, LQCD'13, LQCD'15, and LQCD'22 results in the intermediate and high $q^2$-interval, while in the lower $q^2$ interval, our results are more lower. The overall trend of $f_{\rm T}^{BK}(q^2)$ is consistent compared to the results.
    \item The contribution proportion of the NLO derived from LCSR'08 is 7\%, while our ratio for the total result $f_{+}^{BK}(0)$ and $f_{\rm T}^{BK}(0)$ are 5.8\% and 4\%, respectively. In Table~\ref{Tab:SSEcoefficients}, the quality of fit $\Delta$ of $f_{+;0}^{BK}(q^2)$ and $f_{\rm T}^{BK}(q^2)$ is less than one percent, which relatively guarantees their behavior in large $q^2$-region.
\end{itemize}

\begin{table}[t]
\footnotesize
\centering
\caption{Branching fractions integrated over the total $q^2$-region for dielectron, dimuon, ditau and $\ell$ in which $\ell$ stands for the average of dielectron, dimuon channels. For comparison, the theoretical and experimental results are also presented. (In unit: $10^{-7}$)}
\label{Tab:TotalBR}
\begin{tabular}{l l l l l l l}
\hline
~~~~~~~~~~~~~~~~~~~~~~~ &$\mathcal{B}(B\to K \ell^+ \ell^-)$~~~~~~~~ &$\mathcal{B}(B^+\to K^+ e^+ e^-)$~~~~~~~~& $\mathcal{B}(B^+\to K^+\mu^+\mu^-)$~~~~~~~~& $\mathcal{B}(B^+\to K^+\tau^+\tau^-)$   \\
\hline
This work                               &-                                  &$6.633_{-1.070}^{+1.341}$          &$6.620_{-1.056}^{+1.323}$            &$1.760_{-0.197}^{+0.241}$                 \\
PDG~\cite{ParticleDataGroup:2022pth}    &$4.7\pm0.5$                        &$5.6\pm0.6$                        &$4.53\pm0.35$                       &$<2250$                                  \\
HFLAV~\cite{HFLAV:2022esi}              &$4.63\pm0.19$                      &$5.61\pm0.56$                      &$4.50\pm0.21$                       &-                                         \\
HPQCD'13~\cite{Bouchard:2013mia}        &-                                  &$6.14\pm1.33$                      &$6.12\pm1.32$                       &$1.44\pm0.15$                              \\
HPQCD'22~\cite{Parrott:2022zte}         &7.04(55)                           &7.04(55)                           &7.03(55)                            &1.83(13)                                    \\
LCSR'00~\cite{Ali:1999mm}               &-                                  &-                                  &$5.7_{-1.0}^{+1.6}$                 &$1.3_{-0.17}^{+0.34}$                        \\
LCSR'06~\cite{Wu:2006rd}                &-                                  &$6.11_{-0.72}^{+0.76}$             &$5.18\pm0.92$                       &$1.74_{-0.19}^{+0.2}$                         \\
RQM'02~\cite{Faessler:2002ut}           &-                                  &-                                  &5.5(5.1)                            &1.01(0.81)                                     \\
BHDW'11~\cite{Bobeth:2011nj}            &-                                  &-                                  &-                                   &$1.26_{-0.21}^{+0.40}$                          \\
PQCD'12~\cite{Wang:2012ab}              &$5.50_{-1.18-0.55-0.41}^{+1.59+0.57+0.42}$     &-                      &-                                   &$1.29_{-0.26}^{+0.35}\pm0.08\pm0.11$             \\
MILC'15~\cite{Du:2015tda}               &-                                  &-                                  &$6.0533\pm1.1513$                   &$1.6036\pm0.2206$                                 \\
MNE'18~\cite{Li:2018pag}                &-                                  &-                                  &$5.21\pm0.51$                       &-                                                  \\
BaBar'08~\cite{BaBar:2008jdv}           &-                                  &$5.1_{-1.1}^{+1.2}\pm0.2$          &$4.1_{-1.5}^{+1.6}\pm0.2$           &-                                                 \\
BaBar'12~\cite{BaBar:2012mrf}           &$4.7\pm0.6\pm0.2$                  &-                                  &-                                   &-                                                  \\
BaBar'16~\cite{BaBar:2016wgb}           &-                                  &-                                  &-                                   &$<2250$                                             \\
CDF'11~\cite{CDF:2011buy}               &-                                  &-                                  &$4.6\pm0.4\pm0.2$                   &-                                                    \\
Belle'02~\cite{Belle:2001oey}           &$7.5_{-2.1}^{+2.5}\pm0.9$          &$4.8_{-2.4-1.1}^{+3.2+0.9}$        &$9.9_{-3.2-1.4}^{+4.0+1.3}$         &-                                                     \\
Belle'09~\cite{Belle:2009zue}           &$5.3_{-0.5}^{+0.6}\pm0.3$          &$5.7_{-0.8}^{+0.9}\pm0.3$          &$5.3_{-0.7}^{+0.8}\pm0.3$           &-                                                      \\
Belle'19~\cite{BELLE:2019xld}           &$5.99_{-0.43}^{+0.45}\pm0.14$      &$5.75_{-0.61}^{+0.64}\pm0.15$      &$6.24_{-0.61}^{+0.65}\pm0.16$       &-                                                       \\
LHCb'12~\cite{LHCb:2012juf}             &-                                  &-                                  &$4.36\pm 0.15\pm 0.18$              &-                                                        \\
LHCb'14C~\cite{LHCb:2014cxe}            &-                                  &-                                  &$4.29\pm 0.07\pm 0.21$              &-                                                         \\
LHCb'16~\cite{LHCb:2016due}             &-                                  &-                                  &$4.37\pm0.15\pm0.23$                &-                                                          \\
\hline
\end{tabular}
\end{table}

\begin{table}[t]
\footnotesize
\begin{center}
\caption{Branching fractions integrated over the two various $q^2$-region for dielectron, dimuon. For comparison, the theoretical and experimental results are also presented. (In unit: $10^{-7}$)}
\label{Tab:01BR}
\begin{tabular}{l lllll}
\hline
~~~~~~~~~~~~~~~~~~~~~~~~~~~~~~~~ &$q^2/{\rm GeV^2}$~~~~~~~~~~~~ &$\mathcal{B}(B^+\to K^+ e^+ e^-)$~~~~~~~~~~~~~~& $\mathcal{B}(B^+\to K^+\mu^+\mu^-)$     \\
\hline
This work                            &(1.0,6)                      &$1.873_{-0.307}^{+0.388}$                    &$1.875_{-0.304}^{+0.383}$             \\
This work                            &(1.1,6)                      &$1.834_{-0.301}^{+0.380}$                    &$1.839_{-0.298}^{+0.376}$              \\
\hline
HPQCD'13~\cite{Bouchard:2013mia}     &(1.0,6)                      &-                                            &$1.81\pm0.61$                              \\
HPQCD'22~\cite{Parrott:2022zte}      &(1.0,6)                      &$2.11\pm0.18(\pm0.11)_{{\rm QED}}$           &$2.11\pm0.18(\pm0.04)_{{\rm QED}}$          \\
BHDW'11~\cite{Bobeth:2011nj}         &(1.0,6)                      &-                                            &$1.75_{-0.38}^{+0.64}$                       \\
MILC'15~\cite{Du:2015tda}            &(1.0,6)                      &-                                            &$1.7835\pm0.4251$                             \\
CDF'11~\cite{CDF:2011buy}            &(1.0,6)                      &-                                            &$1.41\pm 0.20\pm 0.10$                         \\
LHCb'12~\cite{LHCb:2012juf}          &(1.0,6)                      &-                                            &$1.205\pm0.085\pm0.070$                         \\
LHCb'14A~\cite{LHCb:2014vgu}         &(1.0,6)                      &$1.56_{-0.15-0.04}^{+0.19+0.06}$             &-                                                \\
Belle'19~\cite{BELLE:2019xld}        &(1.0,6)                      &$1.66_{-0.29}^{+0.32}\pm0.04$                &$2.30_{-0.38}^{+0.41}\pm0.05$                     \\
\hline
HPQCD'22~\cite{Parrott:2022zte}      &(1.1,6)                      &$2.07\pm0.17(\pm0.10)_{{\rm QED}}$           &$2.07\pm0.17(\pm0.04)_{{\rm QED}}$                \\
MILC'15~\cite{Du:2015tda}            &(1.1,6)                      &-                                            &$1.7475\pm0.4251$                                  \\
LHCb'14C~\cite{LHCb:2014cxe}         &(1.1,6)                      &-                                            &$1.186\pm0.034\pm0.059$                             \\
LHCb'19~\cite{LHCb:2019hip}          &(1.1,6)                      &$1.401_{-0.083}^{+0.098}\pm0.069$            &-                                                    \\
LHCb'21~\cite{LHCb:2021trn}          &(1.1,6)                      &$1.401_{-0.069}^{+0.074}\pm0.064$            &-                                                     \\
LHCb'22~\cite{LHCb:2022vje}          &(1.1,6)                      &$1.250_{-0.059}^{+0.064}\pm0.054$            &-                                                      \\
CMS'24~\cite{CMS:2024syx}            &(1.1,6)                      &-                                            &$1.242\pm0.068$                                         \\
\hline
\end{tabular}
\end{center}
\end{table}

\begin{figure}[t]
\begin{center}
\includegraphics[width=0.48\textwidth]{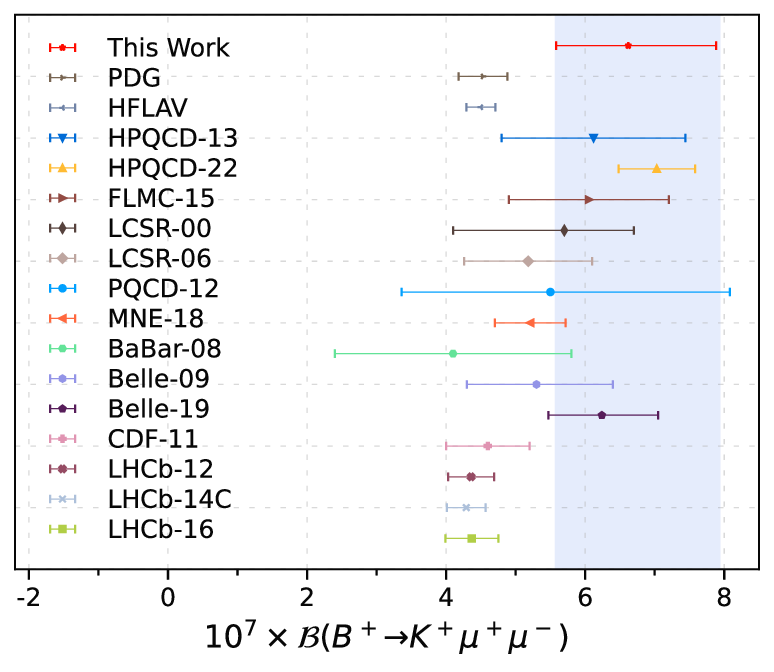}
\end{center}
\caption{Branching fraction of $B^+\to K^+ \mu^+\mu^-$ as a function of the dimuon invariant mass squared ($q^2$) with the shaded band stands for our results, including other results from both the experimentally and theoretically are plotted for comparison.}
\label{Fig:dB_mumu}
\end{figure}
\begin{figure}[t]
\begin{center}
\includegraphics[width=0.48\textwidth]{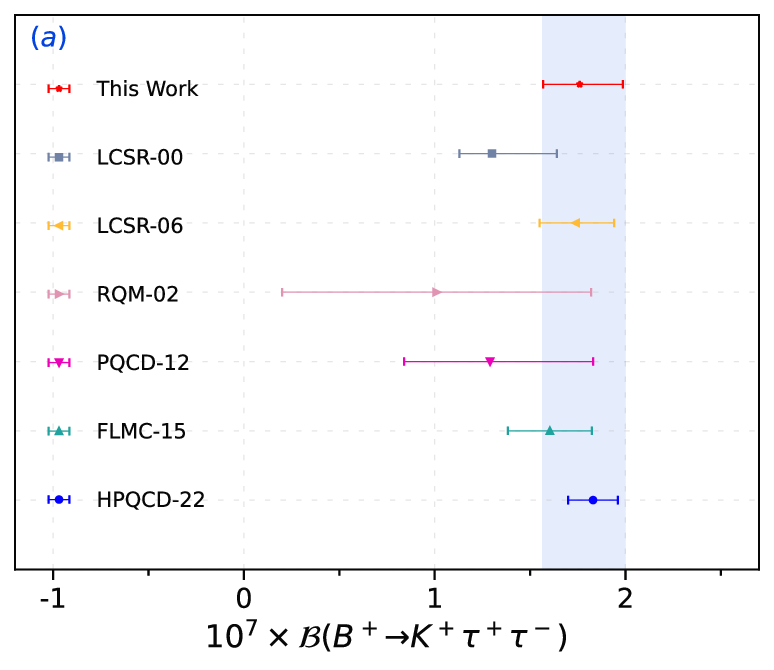}\includegraphics[width=0.48\textwidth]{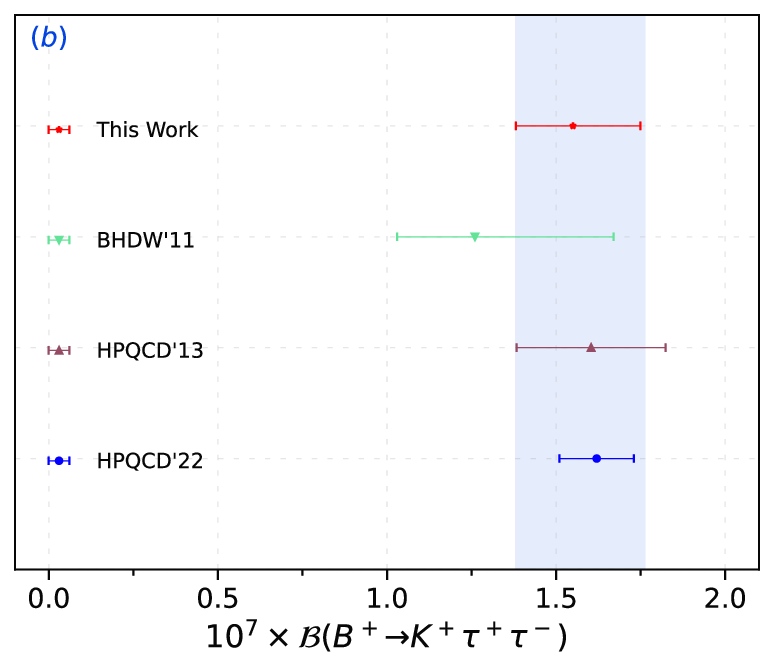}
\end{center}
\caption{Branching fractions of $B^+\to K^+ \tau^+\tau^-$ as a function of the ditau invariant mass squared($q^2$) with the shaded band stands for our results, including other results from both the experimentally and theoretically are plotted for comparison. Planes (a) and (b) stand for the different $q^2$-regions, $\it i.e.$, $(4m_{\tau}^2, q_{\max}^2)~\mathrm{GeV^2}$ and $(14.18, q_{\max}^2)~\mathrm{GeV^2}$, respectively.}
\label{Fig:dB_tautau}
\end{figure}
\begin{figure}[t]
\begin{center}
\includegraphics[width=0.495\textwidth]{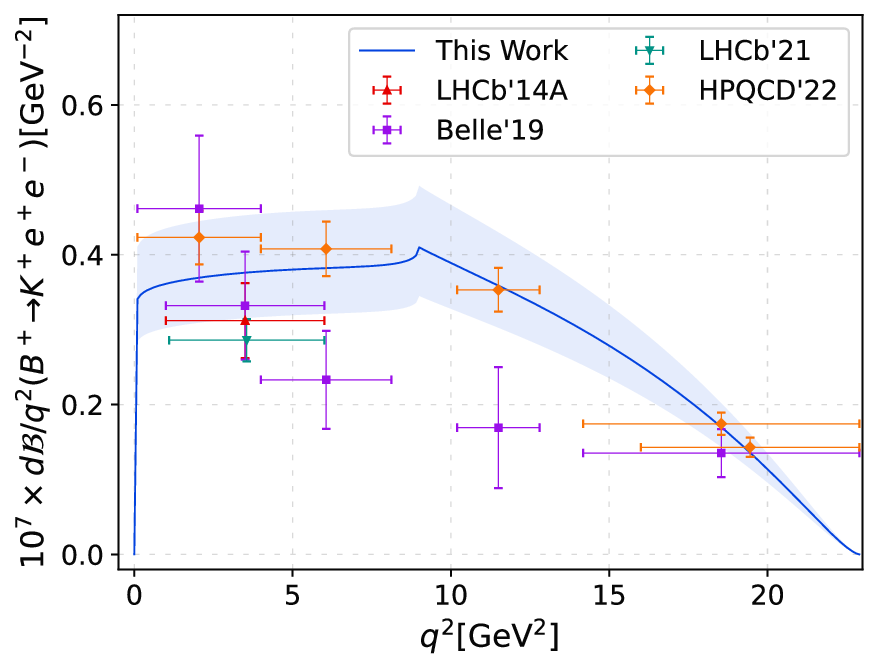}
\end{center}
\caption{Differential branching fraction for $B^+\to K^+ e^+e^-$ as a function of the dielectron invariant mass squared($q^2$) with the shaded bands stand for the uncertainties, including the theoretical and experimental predictions from the LHCb'14A(21)~\cite{LHCb:2014vgu,LHCb:2021trn}, HPQCD'22~\cite{Parrott:2022zte}, and the Belle'19~\cite{BELLE:2019xld} for comparison.}
\label{Fig:ee}
\end{figure}

By means of the obtained $B\to K$ TFFs from the LCSR approach, one can compute the physical SM observables with respect to the rare decays $B\to K\ell^+\ell^-~(\ell=e,\mu,\tau)$. Studying the differential branching fractions helps us understand the internal structure of a particle and its interactions with other particles. According to Eq.~(\ref{Eq:dGamma}), we compute the branching fractions of the rare decays $B\to K\ell^+\ell^-$ in various bins of dilepton invariant mass squared ($q^2$). Then, we exhibit our numerical results in Tables~\ref{Tab:TotalBR} and \ref{Tab:01BR}, including other predictions. Moreover, we also plot the numerical results of the branching fractions for dimuon and ditau channels in Figs.~\ref{Fig:dB_mumu} and \ref{Fig:dB_tautau}, containing other numerical predictions for a more intuitive comparison. In addition, we provide the following statements:
\begin{itemize}
    \item   In Table~\ref{Tab:TotalBR}, we present our predictions that branching fractions integrated over the full $q^2$-region for the dielectron, dimuon, and ditau. For comparison, the theoretical and experimental results also are presented , such as the PDG~\cite{ParticleDataGroup:2022pth}, HFLAV~\cite{HFLAV:2022esi}, HPQCD'13(22)~\cite{Bouchard:2013mia,Parrott:2022zte}, LCSR'00(06)~\cite{Ali:1999mm,Wu:2006rd}, RQM'02~\cite{Faessler:2002ut}, BHDW'11~\cite{Bobeth:2011nj}, PQCD'12~\cite{Wang:2012ab}, MILC'15~\cite{Du:2015tda}, (Majorana neutrino exchange)MNE-18~\cite{Li:2018pag}, BaBar-08(12,16)~\cite{BaBar:2008jdv,BaBar:2012mrf,BaBar:2016wgb}, CDF-11~\cite{CDF:2011buy}, Belle'02(09,19)\cite{Belle:2001oey,Belle:2009zue,BELLE:2019xld}, and LHCb'12(14C,16)~\cite{LHCb:2012juf,LHCb:2014cxe,LHCb:2016due}.

    \item   In Table~\ref{Tab:01BR}, we present the branching fractions in two various $q^2$-region, $\it i.e.,$ $(1.0, 6.0)$ and $(1.1, 6.0)$ $\mathrm{GeV^2}$, including the theoretical and experimental results for comparison. It needs to be mentioned that the LHCb'21's branching fractions with electron channel quoted here are obtained by utilizing the $B\to K\mu^+\mu^-$ results from LHCb'14C, combined with the ratio determined in LHCb'21.

    \item   By observation of the listed results of branching fractions in Tables~\ref{Tab:TotalBR} and \ref{Tab:01BR}, one can conclude that the theoretical predictions of branching fractions are generally larger than the experimental results, except for Belle'19, but their order of magnitude remains the same. Moreover, for the dimuon channel, those results in Table~\ref{Tab:TotalBR} are approximately in the range $2\sim 8\times 10^{-7}$, with their uncertainties also included. Except for RQM'02 and Belle'02, they have a relatively wide margin of error.

    \item   For the ditau channel in Table~\ref{Tab:TotalBR}, our results compare with others, which are in the range $0.2\sim 2\times 10^{-7}$ with good consistency, and both fall inside the upper limit from the predictions of BaBar'16 and PDG~\cite{ParticleDataGroup:2022pth,BaBar:2016wgb}. In which, the results of HPQCD-13 and BHDW-11 begin the integration at $14.18~\mathrm{GeV^2}$.

    \item   In Figs.~\ref{Fig:dB_mumu} and \ref{Fig:dB_tautau}, the former exhibits an intuitive comparison for the branching fractions of dimuon channel based on the results listed in Table~\ref{Tab:TotalBR}, except for RQM'02 and Belle'02, while the latter shows the branching fractions of ditau channel with two different $q^2$-region, $\it i.e.$, $(4m_{\tau}^2, q_{\max}^2)~\mathrm{GeV^2}$ and $(14.18, q_{\max}^2)~\mathrm{GeV^2}$.
\end{itemize}

\begin{figure}[t]
\begin{center}
\includegraphics[width=0.495\textwidth]{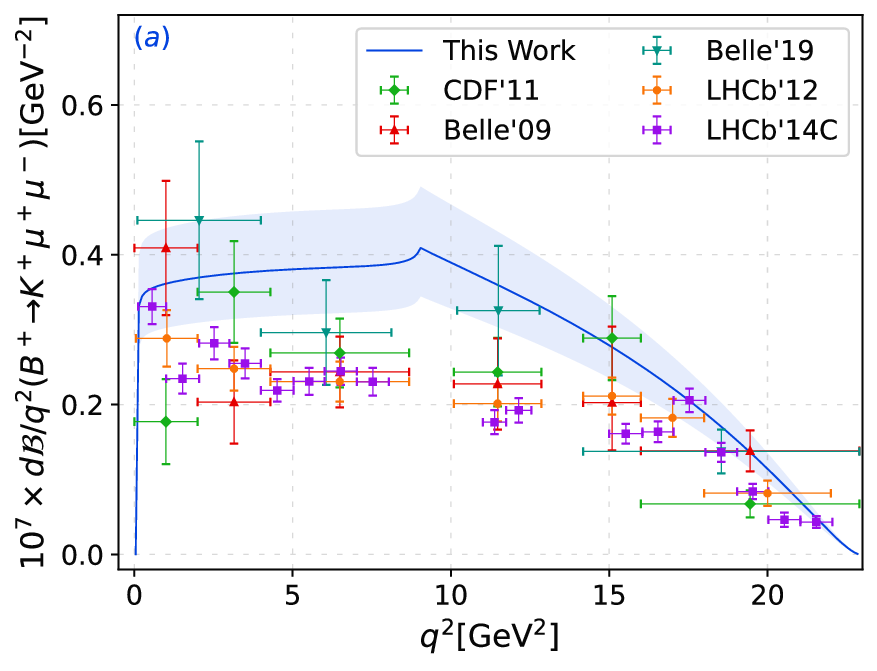}
\includegraphics[width=0.495\textwidth]{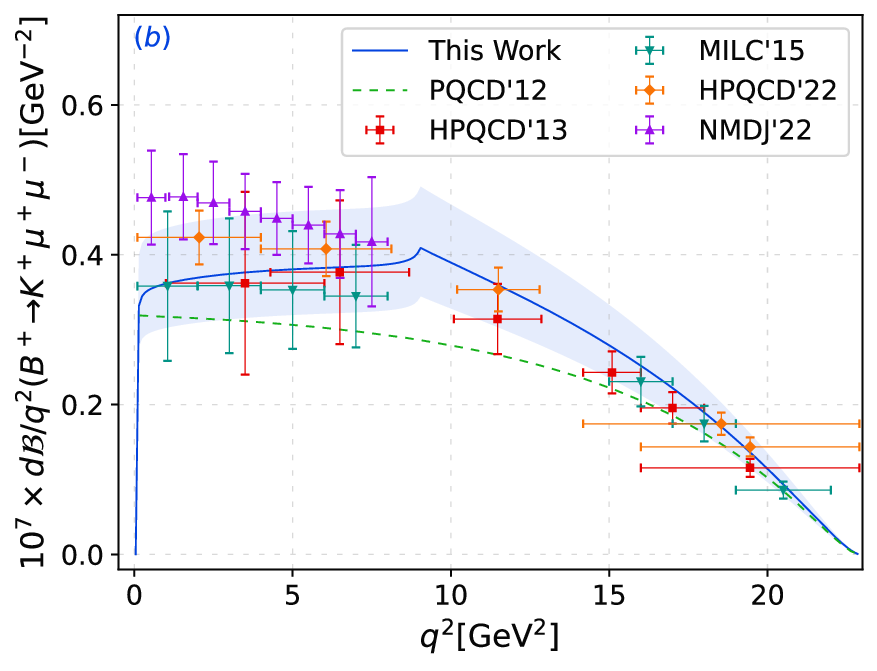}
\end{center}
\caption{Differential branching fraction of $B^+\to K^+ \mu^+\mu^-$ as a function of the dimuon invariant mass squared($q^2$) with the shaded band stands for our results, including other results from both the experimentally and theoretically are plotted for comparison. Plane (a) stands for the comparison of our results with experiments, while plane (b) is the comparison with different theories.}
\label{Fig:dB_mumu_ab}
\end{figure}
\begin{figure}[t]
\begin{center}
\includegraphics[width=0.495\textwidth]{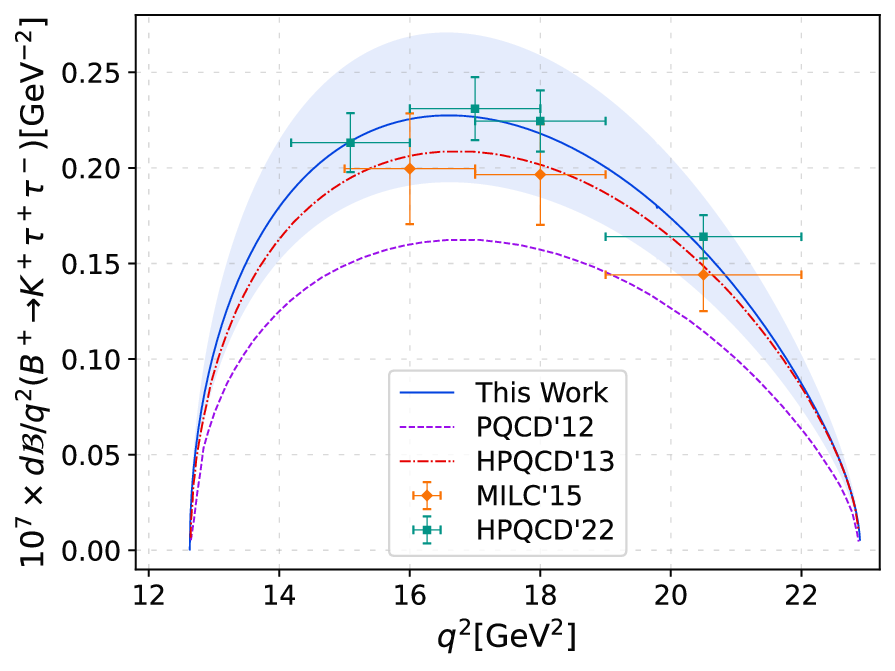}
\end{center}
\caption{Differential branching fraction for $B^+\to K^+ \tau^+\tau^-$ as a function of the ditau invariant mass squared ($q^2$), with the shaded bands stand for the uncertainties, including other results from the HPQCD'13(22)~\cite{Bouchard:2013mia,Parrott:2022zte}, PQCD'12~\cite{Wang:2012ab}, and MILC'15~\cite{Du:2015tda} for comparison.}
\label{Fig:tautau}
\end{figure}

Meanwhile, the behaviors of differential decay widths for $B\to K\ell^+\ell^-$ with $\ell=(e,\mu,\tau)$ are exhibited in Figs.~\ref{Fig:ee}, \ref{Fig:dB_mumu_ab}, and \ref{Fig:tautau}, respectively. In which the solid lines are central values and the shaded bands stand for their uncertainties. Additionally, we provide the following descriptions:
\begin{itemize}
    \item   In Fig.~\ref{Fig:ee}, we present the differential branching fraction of $B^+\to K^+e^+e^-$ in whole physical $q^2$-region. Meanwhile, the theoretical and experimental results also are presented for comparison, $i.e.,$ the LHCb'14A(21)~\cite{LHCb:2014vgu,LHCb:2021trn}, HPQCD'22~\cite{Parrott:2022zte}, and the Belle'19~\cite{BELLE:2019xld}.

    \item   In Fig.~\ref{Fig:dB_mumu_ab}, we present the differential branching fraction of $B^+\to K^+\mu^+\mu^-$ in whole $q^2$-region, compared with the experimental results from the CDF'11~\cite{CDF:2011buy}, the Belle'09~\cite{Belle:2009zue}, Belle'19~\cite{BELLE:2019xld} and the LHCb'12(14C)~\cite{LHCb:2012juf,LHCb:2014cxe}, while the theoretical predictions from the HPQCD'13~\cite{Bouchard:2013mia} and HPQCD'22~\cite{Parrott:2022zte}, PQCD'12~\cite{Wang:2012ab}, MILC'15~\cite{Du:2015tda}, and NMDJ'22~\cite{Gubernari:2022hxn}.

    \item  In Fig.~\ref{Fig:dB_mumu_ab}: plane (a) represents the comparison of our results with experiments, in which our results are in good agreement with the experimental results in the region $14~\mathrm{GeV^2}\leqslant q^2\leqslant q^2_{\mathrm{max}}$, but there are small deviations from the experimental results in the middle and lower regions; plane (b), which is the comparison with different theories, demonstrates that our behaviors are in good accordance with the theoretical predictions, while somewhat lower than the HMDJ'22 in the region $0~\mathrm{GeV^2}\leqslant q^2 \leqslant 3~\mathrm{GeV^2}$. The differences between the theoretical and experimental predictions remain.

    \item The differential branching fractions of $B^+\to K^+\tau^+\tau^-$ with its uncertainties are shown in Fig.~\ref{Fig:tautau}, including others results from the HPQCD'13(22)~\cite{Bouchard:2013mia,Parrott:2022zte}, PQCD'12~\cite{Wang:2012ab}, and MILC'15~\cite{Du:2015tda} for comparison. And our results closely match others, except for PQCD'12.
\end{itemize}

\begin{table}[t]
\footnotesize
\begin{center}
\caption{The ratio of branching fractions with regards to $B\to K\ell^+\ell^-$ in two $q^2$ intervals, i.e., lower-$q^2$ region with $q^2 \in (0.1,1.0)~ {\rm GeV}^2$ and central-$q^2$ region with $q^2\in(1.1,6.0)~{\rm GeV}^2$, compared with the theoretical and experimental results. In which the theoretical predictions are from the Flavio software package.}
\label{Tab:ObsRK}
\begin{tabular}{l l l }
\hline
$\mathcal{R}_K$~~~~~~~~~~~~~~~~~~~~~~~~~~~~ &lower-$q^2$ region~~~~~~~~~~ &central-$q^2$ region   \\
\hline
This work                                &$0.995_{-0.020}^{+0.021}$                   &$1.001_{-0.003}^{+0.003}$                         \\
Flavio~\cite{Straub:2018kue}      &$0.9936\pm0.0003$                           &$1.0007\pm0.0003$                                     \\
\hline
LHCb'14A~\cite{LHCb:2014vgu}             &-                                           &$0.745_{-0.074}^{+0.090}\pm0.036$                       \\
LHCb'19~\cite{LHCb:2019hip}              &-                                           &$0.846_{-0.054-0.014}^{+0.060+0.016}$                    \\
LHCb'21~\cite{LHCb:2021trn}              &-                                           &$0.846_{-0.039-0.012}^{+0.042+0.013}$                     \\
LHCb'22~\cite{LHCb:2022vje}              &$0.994_{-0.082-0.027}^{+0.090+0.029}$       &$0.949_{-0.041-0.022}^{+0.042+0.022}$                      \\
Belle'19~\cite{BELLE:2019xld}            &-                                           &$1.03_{-0.24}^{+0.28}\pm0.01$                               \\
CMS'24~\cite{CMS:2024syx}                &-                                           &$0.78_{-0.23}^{+0.47}$                                       \\
\hline
\end{tabular}
\end{center}
\end{table}

We shall then discuss LU $\mathcal{R}_K$ and flat term $F_{H}^\ell$ with $\ell=\mu,\tau$ in the SM. According to formula~(\ref{Eq:RK}), we calculated $R_K$ in two $q^2$ intervals, $\it i.e.,$ lower-$q^2$: $(0.1,1.1) ~{\rm GeV^2}$ and central-$q^2$: $(1.1,6.0) ~{\rm GeV^2}$. Then, we present our results in Table~\ref{Tab:ObsRK}, including other predictions, which has the following descriptions:
\begin{itemize}
    \item   For comparison, the predictions with different $q^2$ intervals from the SM as calculated by the flavio software package~\cite{Straub:2018kue}, the Belle'19~\cite{BELLE:2019xld}, the CMS'24~\cite{CMS:2024syx}, and the LHCb Collaboration in 2014~\cite{LHCb:2014vgu}, 2019~\cite{LHCb:2019hip}, 2021~\cite{LHCb:2021trn}, and 2022~\cite{LHCb:2022vje} are introduced in Table~\ref{Tab:ObsRK}. Here, we quote results from LHCb'14A and Belle'19 in the range $1~{\rm GeV}^2 \leqslant q^2\leqslant 6~{\rm GeV}^2$.

    \item   Our results with two $q^2$ intervals have agreement with the SM's and experimental predictions, hinting that the LU $\mathcal{R}_K$ maintains a good consistency between theory and experiment and further checking the SM.

    \item   According to the mention of the LHCb Collaboration in 2022~\cite{LHCb:2022vje}, their results differ from previous LHCb measurements, which are superseded. In comparison with the most recent LHCb measurements, our predictions are in good accord with those in the lower-$q^2$ region, while marginally larger than those in the central-$q^2$ region.

    \item   Moving over to theoretical and experimental uncertainties, the statistical uncertainties remain significantly large, and thus it is necessary to continuously reduce the theoretical uncertainties and provide more precise experimental data to keep challenging the SM.
\end{itemize}

\begin{figure}[t]
\begin{center}
\includegraphics[width=0.495\textwidth]{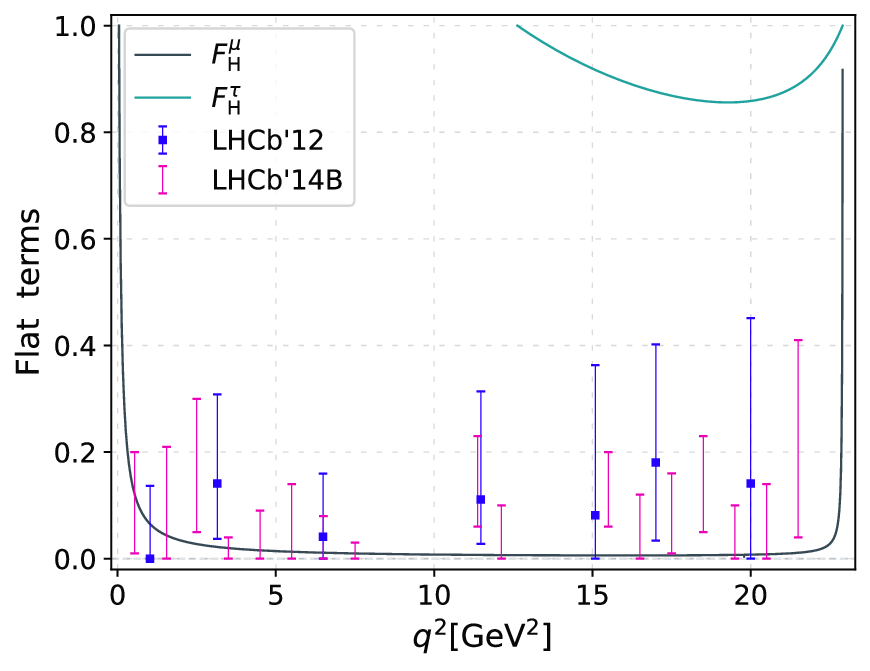}
\end{center}
\caption{Flat terms $F_\mathrm{H}^\ell$ with $\ell=(\mu,\tau)$ for the $B^+\to K^+\ell^+\ell^-$ decay are presented, where the upper/lower curves stand for $\tau$ and $\mu$ channel in the SM, respectively. The results of LHCb'12(14B) come from Refs.~\cite{LHCb:2012juf,LHCb:2014auh}.}
\label{Fig:FH}
\end{figure}

\begin{table}[t]
\footnotesize
\begin{center}
\caption{Branching fraction predictions of the rare decay $B^+\to K^+\nu\bar{\nu}$ are listed, including the experimental and theoretical results. Experimental results are given as $90\%$ confidence limit and marked ``${\rm Exp}$".}
\label{Tab:BRnunu}
\begin{tabular}{l l}
\hline
~~~~~~~~~~~~~~~~~~~~~~~~~~~~~~~~~~~~~~~~~~~~~~~~~~~~~~~~~~~~~ &$10^{6}\times\mathcal{B}(B^+\to K^+\nu\bar{\nu})$                \\
\hline
This work                                         &$4.135_{-0.655}^{+0.820}$                                   \\
Belle'23~\cite{Belle-II:2023esi}                  &$23\pm5({\rm stat})_{-4}^{+5}({\rm syst})$                   \\
Belle'21~\cite{Belle-II:2021rof}                  &$<41(90\% ~{\rm C.L.}){\rm Exp.}$                             \\
Belle'17~\cite{Belle:2017oht}                     &$<16(90\% ~{\rm C.L.}){\rm Exp.}$                              \\
HPQCD'22~\cite{Parrott:2022zte}                   &$5.67(38)$                                                      \\
PQCD'12~\cite{Wang:2012ab}                        &$4.4_{-1.1}^{+1.4}$                                              \\
MILC'15~\cite{Du:2015tda}                         &$4.94(52)$                                                        \\
DB'23~\cite{Becirevic:2023aov}                    &$5.06\pm0.14\pm0.28$                                               \\
BFH'24~\cite{Hou:2024vyw}                         &$4.16\pm0.57$                                                       \\
AJB'21~\cite{Buras:2021nns}                       &$4.53(64)$                                                           \\
AJB'22~\cite{Buras:2022wpw}                       &$4.65(62)$                                                            \\
PSBD'21~\cite{BhupalDev:2021ipu}                  &$3.98(47)$                                                             \\
WA'09~\cite{Altmannshofer:2009ma}                 &$4.50(70)$                                                              \\
JFK'09~\cite{Kamenik:2009kc}                      &$5.10(80)$                                                               \\
\hline
\end{tabular}
\end{center}
\end{table}

Then, we show the flat term $F_\mathrm{H}^\ell$ with $\ell = (\mu, \tau)$ for $B^+\to K^+\ell^+\ell^-$ decay in Fig.~\ref{Fig:FH}, where the flat term $F_\mathrm{H}^\ell$ for $B^+\to K^+\ell^+\ell^-$ as a function of the dimuon and ditau invariant mass squared$(q^2)$ are presented, including the uncertainties but too small to be visible, while the results from the LHCb Collaboration also are provided~\cite{LHCb:2012juf,LHCb:2014auh}. Furthermore, we give the integrated numerical values in the region $q_{\rm{min}}^2 \leqslant q^2 \leqslant q_{\rm{max}}^2$, containing its errors in quadrature:

\begin{eqnarray}
F_\mathrm{H}^{\mu}=0.021_{-0.003}^{+0.003}
\\
F_\mathrm{H}^{\tau}=0.890_{-0.016}^{+0.016}
\end{eqnarray}

At the end of the chapter, we compute the branching fractions of rare decay $B^+\to K^+\nu\bar{\nu}$ as a function of the dineutrino invariant mass squared($q^2$). Then, we have compiled the results from the experimental and theoretical studies and listed them in Table~\ref{Tab:BRnunu}, where those SM predictions are approximately in the $3\sim 6\times 10^{-6}$ range, including the uncertainties. The result of Ref.~\cite{Buras:2021nns} is presented in Table~\ref{Tab:BRnunu} by using CKM $|V_{tb}V_{ts}^*|$. The most recent experimental research on the $B\to K\nu\bar{\nu}$ branching fraction from the Belle Collaboration~\cite{Belle-II:2023esi} gives an explicit boundary, which exists 3.5 standard deviations to the new LQCD results (HPQCD'22~\cite{Parrott:2022zte}), and our prediction can match with them. Combining the theoretical prediction with the data is crucial for perfecting the SM, and we are hoping that more precise theoretical predictions and more experimental data will emerge to continue testing the SM.

\section{Summary}\label{Sec:4}

In this paper, the charmed meson rare decay $B^+\to K^+\ell^+\ell^-(\nu\bar{\nu})$, which involve the FCNC $b\to s\ell^+\ell^-$ and $b\to s\nu\bar{\nu}$ processes, have been studied by means of the QCD sum rules approach. Also, the kaon leading-twist $\phi_{2;K}(x,\mu_0)$ and twist-3 $\phi_{3;p\sigma}^{K}(x,\mu_0)$ LCDAs are briefly introduced. The $B\to K$ TFFs have been calculated by using the QCD LCSR and up to the NLO QCD corrections, while the behaviors of those three TFFs in whole region as a function $q^2$ had presented in Fig.~\ref{Fig:TFFs_Fp_Fz}. For them, we also presented other results in the diagrams for comparison.

Furthermore, the physical observables of the FCNC processes $B^+\to K^+\ell^+\ell^-$ have been discussed by means of the obtained TFFs and LCDAs. Firstly, the $B^+\to K^+\ell^+\ell^-$ differential branching fraction as a function of the dimuon, dielectron and ditau invariant mass squared($q^2$) have been displayed in Figs.~\ref{Fig:ee}, \ref{Fig:dB_mumu_ab}, and \ref{Fig:tautau}, which also contain the experimental and other theoretical predictions for comparison. Meanwhile, for a better understanding of the results between theory and experiment, we have plotted the numerical diagrams and the integrated values of branching fractions in Figs.~\ref{Fig:dB_mumu} (\ref{Fig:dB_tautau}) and Tables~\ref{Tab:TotalBR} (\ref{Tab:01BR}), which contain our results and other theoretical and experimental results, respectively. Secondly, the LU $\mathcal{R}_K$ and the angular distribution `flat term' $F_{\rm H}^\ell$ have been presented, and we have made a detailed comparison with the experimental and SM's predictions for them in Fig.~\ref{Fig:FH} and Table~\ref{Tab:ObsRK}. For $\mathcal{R}_K$, our prediction is compared with the recent LHCb collaboration, which shows a good agreement. Lastly, we have computed the branching fractions with regard to the decay $B^+ \to K^+\nu\bar{\nu}$, which is presented in Table~\ref{Tab:BRnunu}, including the experimental and theoretical results for comparison. \\

\section{Acknowledgments}

Hai-Bing Fu and Tao Zhong would like to thank the Institute of High Energy Physics of Chinese Academy of Sciences for their warm and kind hospitality. This work was supported in part by the National Natural Science Foundation of China under Grant No.12265010, 12265009, 12175025 and No.12347101, the Project of Guizhou Provincial Department of Science and Technology under Grant No.ZK[2023]024, the Graduate Research and Innovation Foundation of Chongqing, China under Grant No.CYB23011 and No.ydstd1912, the Fundamental Research Funds for the Central Universities under Grant No.2020CQJQY-Z003.

\end{document}